\documentstyle[a4,12pt]{article}
\catcode`\@=11

\@addtoreset{equation}{section}
\def\theequation{\arabic{equation}}
\def\theequation{\thesection.\arabic{equation}}

\def\NPB#1#2#3{{\it Nucl.~Phys.} {\bf{B#1}} (19#2) #3}
\def\PLB#1#2#3{{\it Phys.~Lett.} {\bf{B#1}} (19#2) #3}

\def\PRL#1#2#3{{\it Phys.~Rev.~Lett.} {\bf{#1}} (19#2) #3}

\def\JHEP#1#2#3{{\it J. High Energy Phys.} {\bf#1} (19#2) #3}
\newskip\humongous \humongous=0pt plus 1000pt minus 1000pt

\newif\ifdtup

\def\@normalsize{\@setsize\normalsize{15pt}\xiipt\@xiipt
\abovedisplayskip 14pt plus3pt minus3pt%
\belowdisplayskip \abovedisplayskip
\abovedisplayshortskip  \z@ plus3pt%
\belowdisplayshortskip  7pt plus3.5pt minus0pt}
\def\small{\@setsize\small{13.6pt}\xipt\@xipt
\abovedisplayskip 13pt plus3pt minus3pt%
\belowdisplayskip \abovedisplayskip
\abovedisplayshortskip  \z@ plus3pt%
\belowdisplayshortskip  7pt plus3.5pt minus0pt
\def\@listi{\parsep 4.5pt plus 2pt minus 1pt
            \itemsep \parsep
            \topsep 9pt plus 3pt minus 3pt}}

\def\underline#1{\relax\ifmmode\@@underline#1\else
        $\@@underline{\hbox{#1}}$\relax\fi}
\@twosidetrue
\relax

\catcode`@=12

\catcode`\@=11

\def\section{\@startsection{section}{1}{\z@}{3.5ex plus 1ex minus
   .2ex}{2.3ex plus .2ex}{\large\bf}}
\def\thesection{\arabic{section}}

\def\ps@headings{\def\@oddfoot{}\def\@evenfoot{}
\def\@oddhead{\hbox{}\hfill
        \makebox[.5\textwidth]{\raggedright\ignorespaces --\thepage{}--
        \hfill }}
\def\@evenhead{\@oddhead}
\def\subsectionmark##1{\markboth{##1}{}} }

\ps@headings
\catcode`\@=12

\relax

\def\figcap{\section*{Figure Captions\markboth
        {FIGURECAPTIONS}{FIGURECAPTIONS}}\list
        {Fig. \arabic{enumi}:\hfill}{\settowidth\labelwidth{Fig. 999:}
        \leftmargin\labelwidth
        \advance\leftmargin\labelsep\usecounter{enumi}}}
 \relax
\def\tablecap{\section*{Table Captions\markboth
        {TABLECAPTIONS}{TABLECAPTIONS}}\list
        {Table \arabic{enumi}:\hfill}{\settowidth\labelwidth{Table 999:}
        \leftmargin\labelwidth
        \advance\leftmargin\labelsep\usecounter{enumi}}}
 \relax
\def\reflist{\section*{References\markboth
        {REFLIST}{REFLIST}}\list
        {[\arabic{enumi}]\hfill}{\settowidth\labelwidth{[999]}
        \leftmargin\labelwidth
        \advance\leftmargin\labelsep\usecounter{enumi}}}
 \relax

\catcode`\@=11

\def\marginnote#1{}
\newcount\hour
\newcount\minute
\newtoks\amorpm
\hour=\time\divide\hour by60
\minute=\time{\multiply\hour by60 \global\advance\minute by-
\hour}
\edef\standardtime{{\ifnum\hour<12 \global\amorpm={am}%
    \else\global\amorpm={pm}\advance\hour by-12 \fi
    \ifnum\hour=0 \hour=12 \fi
    \number\hour:\ifnum\minute<100\fi\number\minute\the\amorpm}}
\edef\militarytime{\number\hour:\ifnum\minute<100\fi\number\minute}
\def\draftlabel#1{{\@bsphack\if@filesw {\let\thepage\relax
  \xdef\@gtempa{\write\@auxout{\string
    \newlabel{#1}{{\@currentlabel}{\thepage}}}}}\@gtempa
    \if@nobreak \ifvmode\nobreak\fi\fi\fi\@esphack}
     \gdef\@eqnlabel{#1}}
\def\@eqnlabel{}
\def\@vacuum{}
\def\draftmarginnote#1{\marginpar{\raggedright\scriptsize\tt#1}}
\def\draft{\oddsidemargin -.5truein
        \def\@oddfoot{\sl preliminary draft \hfil
        \rm\thepage\hfil\sl\today\quad\militarytime}
        \let\@evenfoot\@oddfoot \overfullrule 3pt
        \let\label=\draftlabel
        \let\marginnote=\draftmarginnote
   
\def\@eqnnum{(\theequation)\rlap{\kern\marginparsep\tt\@eqnlabel}%
\global\let\@eqnlabel\@vacuum}  }
\def\preprint{\twocolumn\sloppy\flushbottom\parindent 1em
        \leftmargini 2em\leftmarginv .5em\leftmarginvi .5em
        \oddsidemargin -.5in    \evensidemargin -.5in
        \columnsep 15mm \footheight 0pt
        \textwidth 250mmin      \topmargin  -.4in
        \headheight 12pt \topskip .4in
        \textheight 175mm
        \footskip 0pt
        
\def\@oddhead{\thepage\hfil\addtocounter{page}{1}\thepage}
        \let\@evenhead\@oddhead \def\@oddfoot{} \def\@evenfoot{}  }
\def\titlepage{\@restonecolfalse\if@twocolumn\@restonecoltrue\onecolumn
     \else \newpage \fi \thispagestyle{empty}\c@page\z@
        \def\thefootnote{\fnsymbol{footnote}} }
\def\endtitlepage{\if@restonecol\twocolumn \else  \fi
        \def\thefootnote{\arabic{footnote}}
        \setcounter{footnote}{0}}  
\catcode`@=12
\relax

\def\ps@headings{\def\@oddfoot{}\def\@evenfoot{}
\def\@oddhead{\hbox{}\hfill
        \makebox[.5\textwidth]{\raggedright\ignorespaces --\thepage{}--
        \hfill }}
\def\@evenhead{\@oddhead}
\def\subsectionmark##1{\markboth{##1}{}} }

\ps@headings

\relax

\def\firstpage#1#2#3#4#5#6{
\begin{document}
\begin{titlepage}
\nopagebreak
\title{\begin{flushright}
        \vspace*{-1.8in}
        {\normalsize LPT-ORSAY 00/43}\\[-9mm]
        {\normalsize LPTM-00/25 }\\[-9mm]
        {\normalsize hep-th/0004165}\\[4mm]
\end{flushright}
\vfill {#3}}
\author{\large #4 \\[1.0cm] #5}
\maketitle
\vskip -7mm     
\nopagebreak 
\begin{abstract} 
{\noindent #6}
\end{abstract}
\vfill
\begin{flushleft}
\rule{16.1cm}{0.2mm}\\[-3mm]
$^{\dagger}${\small Unit{\'e} mixte de recherche du CNRS (UMR 8627).}\\ 
April 2000
\end{flushleft}
\thispagestyle{empty}
\end{titlepage}}

\def\simlt{\stackrel{<}{{}_\sim}}
\def\simgt{\stackrel{>}{{}_\sim}}
\newcommand{\dal}{\raisebox{0.085cm} {\fbox{\rule{0cm}{0.07cm}\,}}}
\newcommand{\dt}{\partial_{\langle T\rangle}}
\newcommand{\dtbar}{\partial_{\langle\overline{T}\rangle}}
\newcommand{\al}{\alpha^{\prime}}
\newcommand{\mst}{M_{\scriptscriptstyle \!S}}
\newcommand{\mpl}{M_{\scriptscriptstyle \!P}}
\newcommand{\dv}{\int{\rm d}^4x\sqrt{g}}
\newcommand{\lv}{\left\langle}
\newcommand{\rv}{\right\rangle}
 \newcommand{\ph}{\varphi}
\newcommand{\abar}{\overline{a}}
\newcommand{\sbar}{\,\overline{\! S}}
\newcommand{\xbar}{\,\overline{\! X}}
\newcommand{\fbar}{\,\overline{\! F}}
\newcommand{\zbar}{\overline{z}}
\newcommand{\dbar}{\,\overline{\!\partial}}
\newcommand{\tbar}{\overline{T}}
\newcommand{\taubar}{\overline{\tau}}
\newcommand{\ubar}{\overline{U}}
\newcommand{\ybar}{\overline{Y}}
\newcommand{\phb}{\overline{\varphi}}
\newcommand{\cm}{Commun.\ Math.\ Phys.~}
\newcommand{\prl}{Phys.\ Rev.\ Lett.~}
\newcommand{\pr}{Phys.\ Rev.\ D~}
\newcommand{\pl}{Phys.\ Lett.\ B~}
\newcommand{\ibar}{\overline{\imath}}
\newcommand{\jbar}{\overline{\jmath}}
\newcommand{\np}{Nucl.\ Phys.\ B~}
\newcommand{\F}{{\cal F}}
\renewcommand{\L}{{\cal L}}
\newcommand{\A}{{\cal A}}
\newcommand{\be}{\begin{equation}}
\newcommand{\ee}{\end{equation}}
\newcommand{\ba}{\begin{eqnarray}}
\newcommand{\ea}{\end{eqnarray}}
\newcommand{\dslash}{{\not\!\partial}}
\newcommand{\gsi}{\,\raisebox{-0.13cm}{$\stackrel{\textstyle >}{\textstyle\sim}$
}\,}
\newcommand{\lsi}{\,\raisebox{-0.13cm}{$\stackrel{\textstyle <}{\textstyle\sim}$
}\,}
\date{}
\firstpage{3118}{IC/95/34} 
{\huge \bf Brane solutions in strings with broken supersymmetry and
dilaton tadpoles} 
{E. Dudas$^{\,a}$ and J. Mourad$^{\,b}$}
{\normalsize\sl
$^a$ LPT$^\dagger$, B{\^a}t. 210, Univ. de Paris-Sud, F-91405 Orsay,
France \\[-3mm]
\normalsize\sl$^b$ LPTM, Site  Neuville III,
Univ. de Cergy-Pontoise, Neuville sur Oise\\[-3mm]
\normalsize\sl F-95031 Cergy-Pontoise, France}
{
The tachyon-free nonsupersymmetric string theories in ten
dimensions have dilaton tadpoles which forbid a Minkowski vacuum.
We determine the maximally symmetric backgrounds for the $USp(32)$
Type I string and the $SO(16)\times SO(16)$ heterotic string.
The static solutions exhibit nine dimensional Poincar\'e symmetry
and have finite 9D Planck and Yang-Mills constants. 
The low energy
geometry is given by a ten dimensional manifold with two boundaries
separated by a finite distance which suggests a spontaneous
compactification of the ten dimensional string theory.
}

\section{Introduction} 
\setcounter{page}{0}

Nonsupersymmetric string models are generically 
plagued by divergences
which raise the question of their quantum consistency. In particular,
ten dimensional nonsupersymmetric models 
have all dilaton tadpoles
and some of them have tachyons in the spectrum.  
It is however believed that some tadpoles do 
not signal an internal
inconsistency of the theory but merely a bakground redefinition \cite{fs}.
In particular, in orientifolds of Type II theories there should be a
difference between tadpoles of Ramond-Ramond (RR)
closed fields and tadpoles of Neveu-Schwarz Neveu-Schwarz (NS-NS) closed
fields. While the first ones cannot be cured by a background
redefinition and  signal an internal inconsistency of the theory asking therefore always
to be cancelled \cite{pc}, the latter ones could in principle be cured
by a background redefinition. The NS-NS tadpoles remove flat directions,
generate potentials for
the corresponding fields (for example dilaton in ten dimensions) and
break supersymmetry. The difference between RR and NS-NS tadpoles
play an important role in (some) orientifold models with broken supersymmetry
recently constructed \cite{sugimoto, ads,bachas}. 

The purpose of this letter is to explicitly find the background of 
the nonsupersymmetric tachyon-free strings in 10D:
the type I model in ten dimensions
\cite{sugimoto}, containing 32 D${\bar 9}$ 
antibranes and 32 O$9_{-}$
planes,
and the heterotic $SO(16)\times SO(16)$\cite{dh,agmv}.
For the two theories, 
there is no background with maximal $SO(10)$ Lorentz
symmetry, a result to be expected by various considerations.
We find explicitly  the classical backgrounds with a 9D Poincar\'e
symmetry. We find an unique solution for the Type I model and two independent solutions
for the heterotic one.
A remarkable  feature (in the Type I and one of the two heterotic backgrounds) of the static solutions is 
that the tenth coordinate is
dynamically compactified in the classical background.
Furthemore, the effective nine-dimensional Planck and Yang-Mills constants
are finite indicating that the low energy physics 
is nine-dimensional.  
Another classical background with maximal symmetry is a
cosmological-type solution which we explicitly exihibit
for the two theories. They both have big-bang type 
curvature singularities.

In section 2 we briefly review the construction of the 
type I $USp(32)$ string. In section 3, we determine its classical
background with maximal symmetry. In section 4, we consider the
$SO(16)\times SO(16)$ heterotic string and finally
we end in section 5 with a discussion of the solutions.  

\section{ The Type I nonsupersymmetric $USp(32)$ string } 

The unique supersymmetric Type I model in ten dimensions is based on
the gauge group $SO(32)$ and contains, by using a modern language (see, for
example, \cite{polchinski}) 32 D9 branes and 32 O$9_{+}$ planes. There
is however another, nonsupersymmetric 
tachyon-free model, with the same closed string
spectrum at tree-level, containing  32 ${\bar D}$9 branes  (i.e. branes of positive tension and
negative RR charge) and 32 O$9_{-}$ planes  (i.e. nondynamical objects with positive tension and
positive RR charge).
  
The open string partition functions are \cite{sugimoto}
\ba
{\cal K}={1 \over 2 ({4 \pi^2 {\alpha'}})^{5}} 
\ \int_0^{\infty} {d \tau_2 \over \tau_2^{6}} 
(V_8-S_8) {1 \over \eta^8} \ , \nonumber \\
{\cal A}= \ \frac{N^2}{2} \ {1 \over ({8 \pi^2 {\alpha'}})^{5}} 
\int_0^{\infty} {d t \over t^{6}} 
 (V_8-S_8)  {1 \over \eta^8} \ , \nonumber \\
{\cal M}= \ \frac{N}{2} \ {1 \over ({8 \pi^2 {\alpha'}})^{5}} 
\int_0^{\infty} {d t \over t^{6}} 
(V_8+S_8) {1 \over \eta^8} \ , \label{1.1}
\ea
where $\alpha' \equiv M_s^{-2}$ is the string tension and
\be
V_8 = { \theta_3^4 - \theta_4^4 \over 2 \eta^4} \quad , \quad 
S_8 = { \theta_2^4 - \theta_1^4 \over 2 \eta^4} \ , \label{1.2}
\ee
where $\theta_i$ are Jacobi functions and $\eta$ the Dedekind function
(see, for example, \cite{polchinski}). In (\ref{1.1}) the various modular
functions are defined on the double covering torus of the corresponding
(Klein, annulus, M\"obius) surface of modular parameter
\be
\tau = 2 i \tau_2 \quad  \quad ({\rm Klein }) \ , \ 
\tau = {i t \over 2}  \quad  \quad ({\rm annulus}) \ , \
\tau = {i t \over 2} +{1 \over 2} \quad  \quad ({\rm Mobius}) , \label{1.3}
\ee
where $\tau = \tau_1 + i \tau_2$ is the modular parameter of the torus
amplitude and $t$ is the (one-loop) open string modulus.

As usual, the annulus and M\"obius amplitudes have the dual
interpretation of one-loop open
string amplitudes and tree-level closed string propagation with the
modulus $l$, related to the open string channel moduli by
\be
{\cal K}: \ l = {1\over 2\tau_2} \ , \ 
{\cal A}: \ l = {2 \over t} \ , \
{\cal M}: \ l = {1 \over 2t} \ . \label{1.4}
\ee
>From the closed string propagation viewpoint, $V_8$ describe the NS-NS sector
(more precisely, the dilaton) and $S_8$ the RR sector, corresponding to
an unphysical 10-form. The tadpole conditions can be derived from the $t
\rightarrow 0$ ($l \rightarrow \infty$) limit of the amplitudes above and read
\be
{\cal K}+{\cal A}+{\cal M} =  \frac{1}{2} \ {1 \over ({8 \pi^2 {\alpha'}})^{5}}
\int_0^{\infty} d l \ \{ (N +32)^2 \times 1 - (N-32)^2 \times 1 \} + \cdots
\ . \label{1.5}
\ee
It is therefore clear that we can set to zero the RR tappole by choosing
$N=32$, but we are forced to live with a dilaton tadpole. The resulting
open spectrum is nonsupersymmetric (the closed spectrum is supersymmetric and given by the Type I supergravity) and contains the vectors of 
the gauge group $USp(32)$ and a fermion in the antisymmetric (reducible) 
representation. However, the spectrum is free of gauge and gravitational
anomalies and therefore the model should be consistent. It is easy to
realize from (\ref{1.1}) that the model
contains 32 ${\bar D}$9 branes  and 32 O$9_{-}$ planes , such that the
total RR charge is zero but NS-NS tadpoles are present, signaling breaking
of supersymmetry in the open sector. The effective action,  identified by writing the 
amplitudes (\ref{1.1}) in the tree-level closed
channel, contains here the bosonic terms
\be
S = {M_s^8 \over 2 } \int d^{10} x \sqrt{-G} e^{-2 \Phi} [ R +4
(\partial \Phi )^2 ] - T_9 \int d^{10} x [ (N+32) \sqrt{-G} e^{-\Phi} 
- (N-32) A_{10} ] + \cdots \ , \label{1.6}
\ee
where $T_9$ is the D9 brane tension and we set to zero the RR two-form
and the gauge fields, which will play no role in our paper.
Notice in (\ref{1.6}) the peculiar couplings of the dilaton and the
10-form to antibranes and O$9_{-}$ planes, in agreement with the general
properties displayed earlier. The RR tadpole $N=32$ is found in
(\ref{1.6}) simply as the classical field equation for the unphysical
10-form $A_{10}$. 

The difference between the supersymmetric $SO(32)$ and nonsupersymmetric
$USp(32)$ model described previously is in the 
M{\"o}bius amplitude describing propagation between (anti)branes and
orientifold planes. Indeed, in the nonsupersymmetric case there is a
sign change in the vector (or NS-NS in the closed channel) character
$V_8$. Both supersymmetric and nonsupersymmetric possibilites are
however consistent with the particle interpretation and factorization of the amplitudes.

As noticed before, the NS-NS tadpoles generate scalar potentials for the
corresponding (closed-string) fields, in our case the (10d) dilaton.
The dilaton potential read
\be
V \sim (N+32) e^{- \Phi} \quad , \label{1.7}
\ee
and in the Einstein basis is proportional to $(N+32) \exp(3
\Phi/2)$. It has therefore the (usual) runaway behaviour towards
zero string coupling, a feature which is of course true in any
perturbative construction. The dilaton tadpole means that the classical
background, around which we must consistently quantize the string, cannot be
the ten dimensional Minkowski vacuum and solutions with lower symmetry
must be searched for. Once the background is corectly identified, there is no NS-NS
tadpole anymore, of course. 
  
\section{The classical background of the nonsupersymmetric $USp(32)$ Type I string}

We are searching for classical solutions of the effective lagrangian (\ref{1.6}) of
the model (\ref{1.1}) in the Einstein frame, which reads
\be
S_E = {1 \over 2k^2} \int d^{10} x \sqrt{-G} [ R - {1 \over 2}
(\partial \Phi )^2 ] - T_9^E \int d^{10} x [ (N+32) \sqrt{-G} e^{ 3 \Phi
\over 2} - (N-32) A_{10} ] + \cdots \ , \label{2.1}
\ee
where $T_9^E$ indicate that the tension here is in the Einstein frame.   
The maximal possible symmetry of the background of the model described
in the previous paragraph has a nine dimensional Poincar\'e isometry
and is of the following form
\be
ds^2=e^{2A(y)}\eta_{\mu\nu}dx^\mu dx^\nu + e^{2B(y)}dy^2 \ , \
\Phi=\Phi(y) \ , \label{2.2} 
\ee  
where $\mu , \nu = 0 \cdots 8$ and the antisymmetric tensor field from
the RR sector, the gauge fields and all fermion fields are set to zero. 
The Einstein and the dilaton field equations with this ansatz are given by
\ba
&&36(A')^2+8A''-8A'B'+{{1}\over{4}}(\Phi')^2 =- \alpha_E \ e^{2B+3\Phi/2}
 \ , \nonumber \\ 
&&36(A')^2-{{1}\over{4}}(\Phi')^2 = - \alpha_E \ e^{2B+3\Phi/2} \ , \nonumber \\
&&\Phi''+(9A'-B')\Phi'= 3 \alpha_E \ e^{2B+3 \Phi/2} \ , \label{2.3} 
\ea
where we defined $\alpha_E = (N+32) k^2 T_9^E = 64 k^2 T_9^E$ and $A' \equiv
dA/dy$, etc. The function $B$ can be gauge-fixed by using the reparametrisation
invariance of the above equations.
It is convenient to choose the coordinate $y$ where
$B=-3\Phi/4$ so that the exponential factors in the equations (\ref{2.3})
disappear. In this coordinate system, the second equation in (\ref{2.3})
is solved in terms of one function\footnote{It can be checked that the other possible
sign choices in (\ref{2.4}) lead to the same solution.} $f$
\be
A' = {1 \over 6} \sqrt{\alpha_E} \ sh f \quad , \quad  
\Phi' = 2 \sqrt{\alpha_E} \ ch f \ . \label{2.4} 
\ee 
The two other field equations become then
\ba
{4 \over 3} \sqrt{\alpha_E} \ f' ch f  + \alpha_E \ e^{2f} &=& - \alpha_E \ ,
\nonumber \\
2 \sqrt{\alpha_E} \ f' sh f  + {3 \over 2} \alpha_E \ e^{2f} &=& {3 \over 2} \alpha_E
\ , \label{2.5}
\ea
and the solution is then
\be
e^{-f} = {3 \over 2} \sqrt{\alpha_E} \ y + c \ , \label{2.6}
\ee  
where $c$ is a constant. By a choice of the $y$ origin and rescaling of
the $x^{\mu}$ coordinates, the final solution in the Einstein frame
reads
\ba
\Phi &=& {3 \over 4} \alpha_E y^2 + {2 \over 3} \ln |\sqrt{\alpha_E} y| + \Phi_0 \ , \nonumber \\
ds_E^2 &=& |\sqrt{\alpha_E} y|^{1/9} e^{- \alpha_E y^2 / 8} \eta_{\mu\nu}dx^\mu dx^\nu 
+|\sqrt{\alpha_E} y|^{-1} e^{-3 \Phi_0 / 2} e^{- 9 \alpha_E y^2 / 8} dy^2 \ . \label{2.7} 
\ea
For physical purposes it is also useful to display the solution in the
string frame, related as usual by a Weyl rescaling $G \rightarrow
e^{\Phi \over 2} G$ to the Einstein frame
\be
A_s = A + {1 \over 4} \Phi \quad , \quad B_s = B + {1 \over 4} \Phi \ . \label{2.8}
\ee
In the string frame the solution reads
\ba
g_s &\equiv& e^{\Phi} = e^{\Phi_0} |\sqrt{\alpha} y|^{2/3} e^{3 \alpha y^2 /4} \ , \nonumber \\
ds^2 &=& |\sqrt{\alpha} y|^{4/9} e^{\Phi_0 / 2} e^{ \alpha y^2 / 4} \eta_{\mu\nu}dx^\mu dx^\nu 
+|\sqrt{\alpha} y|^{-2/3} e^{-\Phi_0 } e^{- 3 \alpha y^2 / 4} dy^2 \ , \label{2.9} 
\ea
where $\alpha = 64 M_s^{-8} T_9$.
The solution (\ref{2.9}) displays two timelike singularities, 
one at the origin $y=0$ and one at infinity $y=+ \infty$, so that the range of the $y$
coordinate is $0<y<+\infty$. The dilaton, on the other
hand, vanishes at $y=0$ and diverges at $y =+ \infty$. 
The brane solution found above (\ref{2.9}) has a striking
feature. Suppose the $y$ coordinate is noncompact $0 < y <
+\infty$. In curved space however, the real radius $R_c$ is given by the
integral
\be  
2 \pi R_c = \int_{0}^{\infty} dy \ e^B =  e^{- \Phi_0 /2} \alpha^{-{1 \over 2}}
\int_0^{\infty} {du \over u^{1/3}} \ e^{-3 u^2/ 8} \ , \label{2.10}
\ee
where $u=\sqrt{\alpha} y$. The result is finite, meaning that despite apparencies the tenth
coordinate is actually compact. The topology of the solution is thus
a ten dimensional manifold with two boundaries at $y=0$ and 
$y=+\infty$ that is $R^9\times S^1/Z_2$.
Moreover, it can be argued that gravity and gauge
fields of the D${\bar 9}$ branes are confined to the nine-dimensional 
noncompact subspace, by computing the nine-dimensional Planck mass and
gauge couplings, respectively
\ba
M_P^7 &=& M_s^8 \int_{0}^{\infty} dy \ e^{7 A_s+B_s-2 \Phi} = M_s^8 \alpha^{-{1\over 2}}
 e^{-3 \Phi_0 /4}
\int_0^{\infty} {du \over u^{1/9}} \ e^{-3 u^2/4} , \nonumber \\
{1 \over g_{YM}^2} &=& M_s^6 \int_{0}^{\infty} dy \ e^{5 A_s+B_s- \Phi}= M_s^6 \alpha^{-{1 \over 2}}
e^{- \Phi_0 /4} \int_0^{\infty} du  \ u^{1/9} e^{- u^2/2} \
. \label{2.11}
\ea
Both of them are finite, which indeed suggest that gravity and gauge
fields of the D${\bar 9}$ branes are confined to the nine-dimensional 
subspace. The relations (\ref{2.10}) and (\ref{2.11}) are in sharp contrast with the
usual flat space relations obtained by compactifying the ten-dimensional theory down to nine-dimensions
on a circle of radius $R_c$
\be
M_P^7 \sim e^{-2 \Phi_0 } R_c M_s^8 \ , \ {1 \over g_{YM}^2} \sim  e^{- \Phi_0 } R_c M_s^6 \ . \label{2.12}
\ee
 By using exactly the same method we find a cosmological
solution for the $USp(32)$ nonsupersymmetric Type I model by searching a
homogeneous  metric of the form
\be
ds^2 = - e^{2B(t)} d t^2 + e^{2A(t)} \delta _{\mu\nu} dx^\mu dx^\nu  \ , \quad
\Phi=\Phi(t) \ , \label{2.13} 
\ee  
where $t$ is a time coordinate. The solution is easily found by following the steps which led to 
(\ref{2.7}) and (\ref{2.9}).
The result in the string frame is
\ba
g_s &=& e^{\Phi} = e^{\Phi_0} |\sqrt{\alpha} t|^{2/3} e^{-3 \alpha t^2 /4} \ , \nonumber \\
ds^2 &=& - |\sqrt{\alpha} t|^{-2/3} e^{-\Phi_0 } e^{3 \alpha t^2 / 4} dt^2 +  
|\sqrt{\alpha} t|^{4/9} e^{\Phi_0 /2} e^{- \alpha t^2 / 4} \delta_{\mu\nu}dx^\mu dx^\nu 
\ . \label{2.15} 
\ea
The metric has a spacelike curvature singularities at $t=0$ and
$t=+\infty$. The laps separating these two singularities, to be interpreted as the real 
time parameter,  
\be
\tau=\int\ dt \ |\sqrt{\alpha} t|^{-1/3} e^{-\Phi_0/2 } e^{3 \alpha t^2 / 8} \ ,
\ee
is infinite. 

\section{Classical background of the $SO(16)\times SO(16)$
heterotic string}

In ten dimensions there is a unique tachyon-free
non-supersymmetric
heterotic string model \cite{dh,agmv}. It can be obtained from 
the two supersymmetric heterotic strings as a  $Z_2$ 
orbifold. The resulting bosonic spectrum
comprises the gravity multiplet: 
graviton, dilaton, antisymmetric tensor and
gauge bosons of the gauge group $SO(16)\times SO(16)$.
Compactifications of this theory were 
considered in \cite{nssw},
and its strong coupling behavior in nine 
dimensions was examined
in \cite{bd}.
It has been shown \cite{agmv} that the cosmological constant 
(the partition function on the torus) at one loop 
is finite and positive, furthemore   
its approximate value is given by 
\be
\Lambda\approx M_s^{10}{{2^6}\over{(2\pi)^{10}}} \times 5.67 \ .
\ee
The effective low energy action for the gravity multiplet is the same as before except for 
the cosmological constant term, which now reads
\be
-\Lambda\int\sqrt{-G}
\ee
in the string metric. The absence of the dilaton in this term reflects the one-loop
nature of the cosmological constant\footnote{A direct computation shows also that the one-loop dilaton tadpole is non-zero and
proportional to the cosmological constant $\Lambda$.}. The same ansatz of the 
Einstein metric
as in the previous paragraph leads, in the Einstein frame, to the equations
\ba
&&36(A')^2+8A''-8A'B'+{{1}\over{4}}(\Phi')^2 =
- \beta_E \ e^{2B+5\Phi/2}
 \ , \nonumber \\ 
&&36(A')^2-{{1}\over{4}}(\Phi')^2 = - \beta_E \ e^{2B+5\Phi/2}, \nonumber \\
&&\Phi''+(9A'-B')\Phi'= 5 \beta_E \ e^{2B+5 \Phi/2} \ , \label{3.3} 
\ea
where we defined $\beta_E = \Lambda_E k^2$.
The gauge which eliminates the exponential factors is now
$B=-5\Phi/4$. After solving the second equation in (\ref{3.3})
$A'=\sqrt{\beta_E}sh (h)/6$, $\Phi'=2\sqrt{\beta_E} ch(h)$,
the remaining equations give
\be
e^{h}={{1}\over{2}}{{e^{\sqrt{\beta_E}y}+\epsilon 
e^{-\sqrt{\beta_E} y}}\over{e^{\sqrt{\beta_E}y}-\epsilon 
e^{-\sqrt{\beta_E} y}}},
\ee
where $\epsilon =\pm 1$. An important difference with respect
to the type I solution is that here we have two non-equivalent
(that is, not related by coordinate transformations) 
solutions corresponding to $\epsilon=1$ or $-1$.
Let us first consider the $\epsilon=1$ case. 
The solution in the Einstein frame reads
\ba
\Phi &=& \Phi_0+{{1}\over{2}}\ln{ |sh\sqrt{\beta_E}y| }+
2\ln{(ch\sqrt{\beta_E}y)} \ , \nonumber \\
ds^2_E \!\!\!\!&=&\!\!\!\! | sh(\sqrt{\beta_E}y)|^{1 \over 12}
\left(ch(\sqrt{\beta_E}y)\right)^{-1 \over 3}\!\! dx^2
\!\!+\!e^{-5\Phi_0/2} | sh(\sqrt{\beta_E}y) |^{-5 \over 4}
\left(ch(\sqrt{\beta_E}y)\right)^{-5}\!\! dy^2 \label{3.4} 
\ea

and in the string frame
\ba
e^{2\Phi} &=& e^{2\Phi_0} |sh(\sqrt{\beta}y) |
\left(ch(\sqrt{\beta}y)
\right)^4 \ , \nonumber \\
ds^2 &\!\!=\!\!& e^{\Phi_0/2} | sh(\sqrt{\beta}y) |^{1 \over 3}
\left(ch(\sqrt{\beta}y)
\right)^{2 \over 3}
dx^2\!+\! e^{-2\Phi_0} | sh(\sqrt{\beta}y) |^{-1}
\left(ch(\sqrt{\beta}y)\right)^{-4}dy^2 \ , \label{3.5}
\ea
where here $\beta=\Lambda M_s^{-8}$.
In both metrics, the solution has two timelike singularities at
$y=0$ and $y=\infty$.
These singularities are separated by a finite distance 
which in the string frame reads
\be 
2\pi R_c=(\beta)^{-1/2}e^{-\Phi_0}
\int_{0}^{+\infty} du \ (sh \ u)^{-1/2}(ch \ u)^{-2} \ . \label{3.6}
\ee
The spacetime has therefore the topology of a nine dimensional 
Minkowski space times an interval.
Notice that the nine dimensional  Planck mass
\be
M_p^7=M_s^8(\beta)^{-1/2} e^{-5\Phi_0/4}\int_{0}^{\infty}\ du
\left(sh \ u \right)^{-1/3}\left(ch \ u \right)^{-11/3} \ , \label{3.7}
\ee
as well as the 9D Yang-Mills coupling
\be
{{1}\over{g_{YM}^2}}=M_s^6e^{-7\Phi_0/4}(\beta)^{-1/2}\int_{0}^{\infty}du
\left(sh \ u \right)^{-2/3}\left(ch \ u \right)^{-13/3} \ , \label{3.8}
\ee
 are finite. This fact together with the finitude of the length
 of the tenth coordinate indicate that the low energy
 processes are described by a 9D theory.
 The coupling constant vanishes at $0$ and becomes infinite at
large $y$. We shall comment more on this fact in the conclusion.

The second solution corresponding to $\epsilon=-1$
can be obtained  by
exchanging $(ch(\sqrt{\beta}y)$ 
with  $|sh(\sqrt{\beta}y|$ in 
the solutions  (\ref{3.4}) and  (\ref{3.5}).
This solution has two singularities at $0$ and $\infty$, however
the nine-dimensional Planck and Yang-Mills constants as well as
the 
length of the tenth coordinate are infinite.

The cosmological solution invariant with respect to the nine
dimensional Euclidian group can be also readily found
and it reads in the string  frame
\ba
ds^2 &=& -  e^{-2 \Phi_0 } (\sin{\sqrt{\beta}t})^{-1}
(\cos{\sqrt{\beta}t})^{-4}dt^2 +  e^{ \Phi_0/2}
(\sin{\sqrt{\beta}t})^{1/3}(\cos{\sqrt{\beta}t})^{2/3}dx^2 \ , \nonumber \\
e^{2\Phi} &=& e^{2\Phi_0}(\sin{\sqrt{\beta}t})(\cos{\sqrt{\beta}t})^{4} \ .
\ea
The variable $t$ in these equations belongs to the
interval $[0,\pi/(2\sqrt{\beta})]$.
At the boundaries in $t=0$ and $t=\pi/(2\sqrt{\beta})$
the metric develops curvature singularities. 
The time separating these two
singularities is infinite:
\be 
\tau=\int_{0}^{\pi/(2\sqrt{\beta})} dt \ 
(\sin{\sqrt{\beta}t})^{-1/2} (\cos{\sqrt{\beta}t})^{-2}=\infty.
\ee
Notice that the solution obtained by exchanging the sine and
cosine in the above
equations is not a new one since it can be obtained by shiting
the time coordinate $t\rightarrow \pi/(2\sqrt{\beta})-t$. 

\section{Discussion}

Before discussing the solutions we have found
we should mention that there exists an another interesting
non supersymmetric and tachyon-free model
in ten dimensions \cite{sa}: it is an orientifold of the 
type 0B string with a projection that removes the tachyon
and introduces an open sector with the gauge group $U(32)$.
This model has both a one loop (positive) cosmological
constant and a disk dilaton tadpole so 
we have a sum of two terms
in the
low energy effective action
\be
-\Lambda_1\int\sqrt{-G}e^{-\Phi}-\Lambda_2\int\sqrt{-G},
\ee
in the string metric. The first term is of the type we encountered
in the $USp(32)$ case and the second is analogous to the one we
met in the $SO(16)\times SO(16)$ case.
There is no simple gauge choice for $B$ that renders 
the equations as simple as before. 
However qualitatively the solution 
should behave as the $USp(32)$ case for small $y$
where the coupling constant is small and the behavior for large
$y$
should resemble that of the $SO(16)\times SO(16)$ case.
In partcular we expect the radius of the tenth dimension to be
compactified (since the divergence of the radius in the second
solution of $SO(16)\times SO(16)$ is due to the
behavior at the origin).
 
We have determined the maximally symmetric 
solutions to the low energy equations 
of  two tachyon-free non-supersymmetric strings.
To what extent can we consider these solutions as 
representing the  vacuum of these string theories ?
Perturbatively, there are two kind of string corrections
to the low energy effective action.
The first ones are $\alpha'$ corrections involving the string
oscillators and the second ones 
are $g_s$ string loop corrections.
A common feature of the solutions we found is that the conformal
factor $e^{2A}$ (in nine dimensions) as 
well as the string coupling
vanish at the origin and diverge at $y=\infty$.
The effective string scale at coordinate $y$ being given by 
$M_s^2(y)=M_s^2e^{2A_s}$, we expect $\alpha'$ corrections to be
important at the origin and the loop corrections to
be dominant at infinity. So strictly speaking 
we cannot trust the classical solution near the 
two singularities
where interesting string physics would occur.
Another less ambitious question, 
is the classical stability of
our solutions. That is, do small perturbations around the
background we found destroy the solution ? The answer to this
question is closely related to the determination of the
Kaluza-Klein excitations \cite{dm}. 

Another common feature of the static solution of the Type I model (\ref{2.9}) 
and of the first heterotic solution (\ref{3.5}) is that the effective Yang-Mills 
and Planck constants are finite which means that the gravitational and gauge 
physics is effectively nine-dimensional. 
A remarkable feature of the static solutions 
in the low energy approximation is 
the spontaneous compactification  of one coordinate.
Whether this feature will survive the string and loop
corrections is an interesting and open question.
The fact that the
nine dimensional metric is flat in spite of the 
ten-dimensional cosmological constant is in the spirit
of the higher dimensional mechanisms 
which try to explain the vanishing of the (effective)
cosmological constant \cite{cc}. As in the new approaches to this problem discussed 
in \cite{cc}, however, a better understanding of the naked singularities present in our solutions 
is needed in order to substantiate this claim.

A notable difference between the type I and heterotic
solutions is that the type I background is unique, 
whereas there are two classical heterotic backgrounds
with physically very different properties. 
It is possible that this is due to the low energy approximation
and that this degeneracy will be lifted by string corrections.

All of  the nontrivial features of the solutions are due to the presence of 
dilaton tadpoles. The latter are generic for non-supersymmetric
string models. According to the Fischler-Susskind mechanism,
the quantization around the classical solution should lead
to finite string amplitudes \cite{fs,polchinski}. It would be intersting to confirm
 this explicitly for the present models.

\vskip 16pt
\begin{flushleft}
{\large \bf Acknowledgments}
\end{flushleft}

\noindent We grateful to  C. Angelantonj and A. Sagnotti for 
illuminating discussions on nonsupersymmetric strings and to C. Grojean and S. Lavignac for discussions
concerning naked singularities in relation with the proposals \cite{cc}. E.D. would like to thank the Theory
Group at LBNL--Berkeley for warm hospitality during the final stage of this work.  

\end{document}